\documentclass[review]{elsarticle}

\biboptions{sort&compress} 

\journal{Carbon}

\usepackage{bm}
\usepackage{graphicx}
\usepackage{color}
\usepackage{amssymb}
\usepackage{siunitx}
\usepackage{lineno,hyperref}

\newcommand{\vc}[1]{\bm{#1}}

\begin{document}

\begin{frontmatter}

\title{Lattice thermal conductivity of graphene nanostructures}

\author[complutense,warwick]{M. Saiz-Bret\'{\i}n\corref{correspondingauthor}}
\cortext[correspondingauthor]{Corresponding author}
\ead{marta.saiz.bretin@ucm.es}

\author[complutense]{A. V. Malyshev}
\author[complutense]{F. Dom\'{\i}nguez-Adame}
\author[warwick]{D. Quigley}
\author[warwick]{R. A. R\"{o}mer}

\address[complutense]{GISC, Departamento de F\'{\i}sica de Materiales, Universidad Complutense, E-28040 Madrid, Spain}
\address[warwick]{Department of Physics, University of Warwick, Coventry, CV4 7AL, United Kingdom}

\begin{abstract}

Non-equilibrium molecular dynamics is used to investigate the heat current due to the atomic lattice vibrations in graphene \emph{nanoribbons} and \emph{nanorings} under a thermal gradient. We consider a wide range of temperature, nanoribbon widths up to $\SI{6}{\nano\metre}$ and the effect of moderate edge disorder. We find that narrow graphene nanorings can efficiently suppress the lattice thermal conductivity at low temperatures ($\sim\SI{100}{\kelvin}$), as compared to nanoribbons of the same width. Remarkably, rough edges do not appear to have a large impact on lattice energy transport through graphene nanorings while nanoribbons seem more affected by imperfections. Furthermore, we demonstrate that the effects of hydrogen-saturated edges can be neglected in these graphene nanostructures.

\end{abstract}


\end{frontmatter}


\section{Introduction} 

Advances in nanotechnology demand a better understanding of heat transport in nanoscale systems. The increased levels of dissipated power in ever smaller devices make the search for high thermal conductors essential~\cite{Pop06,Vasileska08,Pop10}. On the other hand, thermoelectric energy conversion requires materials with a strongly suppressed thermal conductivity, but still high electronic conduction. In this regard, one of the main goals in thermoelectric research is to find materials with a high figure of merit $ZT=S^2\sigma T / \kappa$, which  assesses the thermoelectric efficiency of a system~\cite{Goldsmid10}. Here $S$ stands for the Seebeck coefficient, and $\sigma$ and $\kappa$ are the electric and thermal conductivities at a temperature $T$, respectively~\cite{Villagonzalo1999}. Both electrons and lattice vibrations contribute to the heat current and consequently $\kappa=\kappa_\mathrm{el} + \kappa_\mathrm{lat}$. Therefore, it is necessary to minimize both contributions to $\kappa$ while keeping $\sigma$ and $S$ high. Unfortunately, parameters $\kappa$, $\sigma$ and $S$ cannot be adjusted independently in most bulk materials. For instance, the ratio $\sigma/\kappa_\mathrm{el}$ in metals is determined from the Wiedemann-Franz law~\cite{Franz1853}. Hence reducing the lattice thermal conductivity $\kappa_\mathrm{lat}$ by increasing phonon scattering is one of the most promising routes to improve thermoelectric materials.

Several works have demonstrated theoretically~\cite{Hicks93,Khitun00,Balandin03,Sadeghi15} and experimentally~\cite{Venkata01,Harman02,Hochbaum08,Boukai08} that nanometer-sized objects exhibit thermoelectric efficiency unachievable with bulk materials. In particular, quantum effects allow thermoelectric devices to overcome the limitations arising from the classical Wiedemann-Franz law: nanodevices with sharp resonances in the electron transmission (such as Fano lineshapes) are in principle ideal candidates for highly efficient waste heat-to-electricity converters because the ratio $\sigma/\kappa_\mathrm{el}$ increases well above the Wiedemann-Franz limit~\cite{Mahan96,GomezSilva12,Zheng12,Garcia13,Fu15,SaizBretin16,Wang16}. Thus, ballistic electrons in nanodevices pave a possible way to achieve large $ZT$ and consequently more efficient thermoelectric devices as refrigerators and generators~\cite{Koumoto13}.

Graphene nanoribbons (GNRs) can behave as single-channel room-tem\-per\-a\-ture ballistic electrical conductors on a length scale greater than ten microns~\cite{Baringhaus14}. Recent advances in nanotechnology enable the fabrication of devices based on GNRs, such as quantum nanorings~\cite{Russo08,Smirnov12,Schelter12,Cabosart14,Samal15}, that can show ballistic transport and consequently are good candidates to exploit quantum effects even at room temperature. Although ballistic electron transport yields higher values of both $\sigma$ and $\kappa_\mathrm{el}$, it turns out that the ratio $S^2\sigma/\kappa_\mathrm{el}$ becomes largely enhanced in quantum nanorings due to the occurrence of Fano anti-resonances~\cite{SaizBretin15}. However, graphene occupies a unique place amongst materials in terms of its thermal properties because it possesses one of the highest lattice thermal conductivities. A high value of $\kappa_\mathrm{lat}$ is undesirable for thermoelectric applications but it can be greatly reduced in GNRs by rough edges~\cite{Savin10}, hydrogen-passivation~\cite{Hu10} and patterning ~\cite{Mazzamuto11,Li14,Zhang12,Chen10,Xu10}. Since graphene is envisioned as a material of choice for a variety of applications in future electronics, understanding how heat is carried in different graphene nanostructures is crucial. Among these structures, graphene nanorings stand out because of the straightforward way in which they exploit quantum interference effects. These effects could be used for designing new quantum interferometers~\cite{Wu10,Munarriz11,Mrenka16,Sousa17} or spintronic devices \cite{Munarriz12,Farghadan13}. Recently, we demonstrated theoretically that graphene nanorings might be useful as thermoelectric devices too~\cite{SaizBretin15}. We found that quantum interference effects lead to large $S$ and hence high $ZT$ when the heat current is solely due to electrons. Yet, lattice heat conduction, which is expected to be the most important contribution to heat transport in carbon materials due to the strong covalent $sp^{2}$ bonding, had not been studied in graphene nanorings.

In this work, we address the contribution of the atomic lattice to heat transport in armchair GNRs and nanorings by using non-equilibrium molecular dynamics (NEMD) simulations as implemented in the LAMMPS Molecular Dynamics Simulator~\cite{LAMMPS}. NEMD simulations provide a direct method to calculate the lattice thermal conductivity. To this end, a heat flow through the system under study establishes a temperature gradient across the system and Fourier's law brings about an estimate of the lattice thermal conductivity. We compare the thermal conductivities of GNRs and rectangular graphene nanorings of widths up to $\SI{6}{\nano\meter}$ over a wide range of temperature. Our study proves that the lattice thermal conductivity $\kappa_\mathrm{lat}$ is greatly reduced in nanorings as compared to GNRs due to higher scattering of lattice vibrations at bends. We also demonstrate that edge disorder has a weaker impact on the heat current in nanorings as compared to GNRs. Similarly, we find that the effects of hydrogen-saturated edges can be safely neglected in these nanostructures.

\section{Model and methodology}

\begin{figure}[tb]
\begin{center}
   \includegraphics[width=0.7\linewidth]{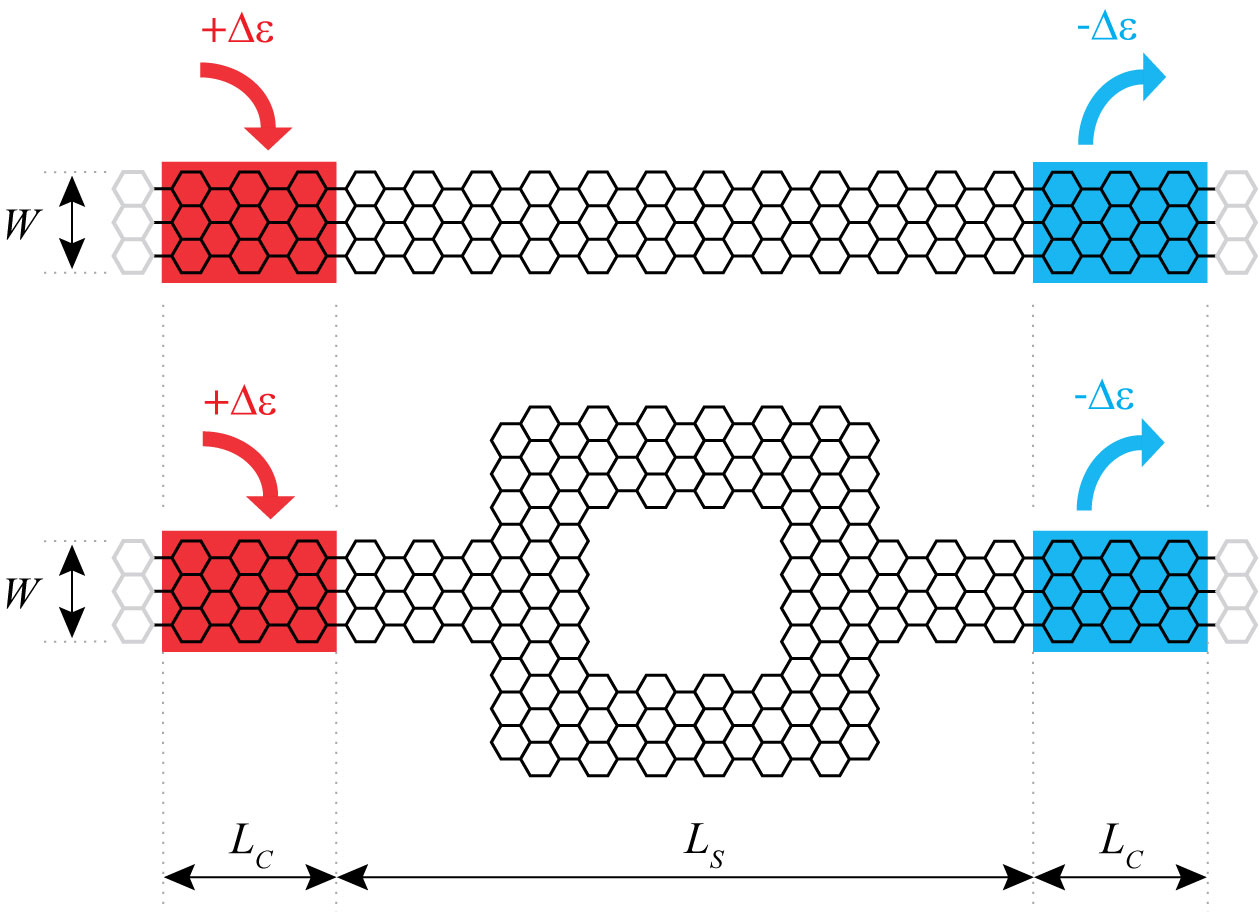}
\end{center}
\caption{Schematic view of the analysed structures. The top panel shows a GNR of width $W$. The bottom panel displays a "square" graphene nanoring. The red (blue) area represents the hot (cold) contact where an amount of heat $\Delta \epsilon$ is introduced (removed) in every time step of the NEMD simulation.}
\label{fig1}
\end{figure}

In our study we focus on two different types of graphene nanostructures connected to leads. The first system under consideration is a uniform rectangular armchair GNR of width $W$, as seen in the top panel of Figure~\ref{fig1}.  We only consider armchair GNRs since many studies have shown them to have a lower thermal conductivity than zig-zag GNRs \cite{Guo09,Hu09,Evans10}, thus being more suitable for thermoelectric applications. The second kind of nanostructures are graphene nanorings. For constructing the latter, a rectangular graphene ring is inserted between two armchair nanoribbons of width $W$, as depicted in the lower panel of Figure~\ref{fig1}. The existence of bends along with the presence of two types of edges (zigzag and armchair) give rise to stronger scattering of lattice vibrations which is expected to reduce the lattice thermal conductivity, as compared with uniform GNRs. We will show later that this is actually the case. 

The lattice thermal conductivity is calculated using NEMD simulations~\cite{LAMMPS,Plimpton95} in which the classical trajectories of Lagrangian particles (in our case, carbon atoms) are obtained by numerically solving Newton's equations of motion. Atoms with initial positions and velocities are exposed to collisions governed by an empirical interatomic potential. At each time step, the force acting on each atom is obtained, and then, positions and velocities are updated.This is an excellent approximation for a wide range of materials; only when we consider light atoms or vibrational motion with a frequency $\omega$ such that $\hbar\omega > k_B T$ should we worry about quantum effects \cite{Frenkel96}. As a general rule, the classical treatment will be valid if the interparticle distance is much larger than the thermal de Broglie wavelength $\Lambda_\mathrm{th}=h/\sqrt{2\pi m k_B T}$. This is indeed the scenario in our simulations since the distance between C atoms is about $\SI{0.142}{\nano\metre}$ while $\Lambda_\mathrm{th}=\SI{0.030}{\nano\metre}$ at room temperature and $\Lambda_\mathrm{th}=\SI{0.043}{\nano\metre}$ at $T=\SI{100}{\kelvin}$.

NEMD simulations provide a direct method to calculate lattice thermal conductivity by applying a perturbation to the system and measuring the response. Most suitable choices of perturbations would be either imposing a thermal gradient $\nabla T$ across the system or introducing a heat flow $\vc{J}$. Throughout this work we consider the latter case, that is, a heat flow is introduced and the subsequent thermal gradient is calculated (further details about this choice will be given in next section)\cite{Lukes00,Pei11}. Then, Fourier's law is applied to obtain the lattice thermal conductivity $\kappa$ (because we only address the lattice contribution to the thermal conductivity we omit the subscript hereafter, unless otherwise stated)
\begin{equation}
\vc{J}=\kappa\, \nabla T\ .
\label{fourier}
\end{equation}
The small size of the systems under study poses a question about the validity of Fourier's law at the nanoscale but a comparison with
more elaborated approaches, such as the phonon Boltzmann transport equation, is beyond the scope of the present manuscript. Nevertheless, previous studies on steady-state thermal transport in nanostructures concluded that Fourier's law is essentially exact in the diffusive and ballistic limits (see Ref.~\cite{Kaiser17} and references therein for further details).

In this work, a time step of $\SI{0.5}{\femto\second}$ is used in the simulations and carbon-carbon interactions are described by the adaptive intermolecular reactive bond order (AIREBO) potential~\cite{Stuart00}, which depends not only on the distance between atoms but also on the local atomic environment, and therefore implicitly contains many-body information. This potential has already been successfully implemented to study thermal and mechanical properties of graphene \cite{Ng12,Pei11,Yang13}. The initial configuration is first equilibrated at a temperature $T$ (typically the mean target temperature) during $\num{2}\times\num{10}^6$ time steps by keeping the two outermost rows of atoms at each end fixed (gray lines in Figure~\ref{fig1}) while applying a Nose-Hoover thermostat to the rest of the atoms which are free to move in three dimensions. Then, we introduce a heat flow by adding at each time step a small amount of energy ($+\Delta \varepsilon$) into the hot contact and removing the same amount of energy ($-\Delta \varepsilon$) from the cold contact. This energy addition (subtraction) is done by rescaling the velocity vectors at both contacts. In order to avoid non-linear temperature profiles when $\Delta \varepsilon$ is too large or temperature fluctuations when $\Delta \varepsilon$ is too small, we adjust $\Delta \varepsilon$ so that $\Delta T = T_\mathrm{max}-T_\mathrm{min}\simeq 0.2 T$. Since the value of $\Delta \varepsilon$ is unknown beforehand it is found by performing iterative simulations and adjusting $\Delta \varepsilon$ at each iteration step to obtain the target $\Delta T/T$ ratio. The system is then switched to the constant volume and constant energy ensemble and we run at least $\num{10}^{7}$ time steps to allow the system to attain steady state. Once it is reached, the three components of velocity are averaged over $10^7$ time steps. Then, the system is divided into slices and the temperature within each slice is obtained from the expression $T=(M/3Nk_B)\sum_i \Big(\langle v_i \rangle_x^2 + \langle v_i \rangle_y^2 + \langle v_i \rangle_z^2\Big)$, where $M$ is the mass of the carbon atom, $N$ is the number of atoms in each slab, $k_B$ denotes the Boltzmann's constant and $\langle v_i \rangle_\mu$ stands for the time averaged $\mu$-component of velocity of atom $i$. The temperature gradient $\partial T/\partial x$ is determined after a linear fit and the heat current, which can be defined as the amount of energy transferred per unit time and cross sectional area, is calculated as
\begin{equation}
J=\frac{1}{Wd}\,\frac{\Delta \epsilon}{\Delta t}\ ,
\end{equation}
where $\Delta t$ is the time step and $d$ the graphene thickness, taken approximately as the diameter of a carbon atom $d=\SI{0.144}{\nano\meter}$ \cite{Guo09}. Finally, the lattice thermal conductivity is obtained using Fourier's law (equation \ref{fourier}).

We have also tried an alternative method of the conductivity calculation, consisting in maintaining a fixed temperature difference between the contacts and calculating the energy flux. In this case temperature profiles can be highly nonlinear, characterized by abrupt temperature changes at the contacts (resulting from a mismatch between the dispersion relation of the fixed temperature parts and the rest of the system~\cite{Shiomi14}). To avoid the non-physical temperature kinks, larger contacts have to be considered increasing the simulation time considerably, for which reason the latter approach was abandoned.

\section{Influence of the contacts}

\begin{figure}[tb]
\begin{center}
   \includegraphics[width=0.7\linewidth]{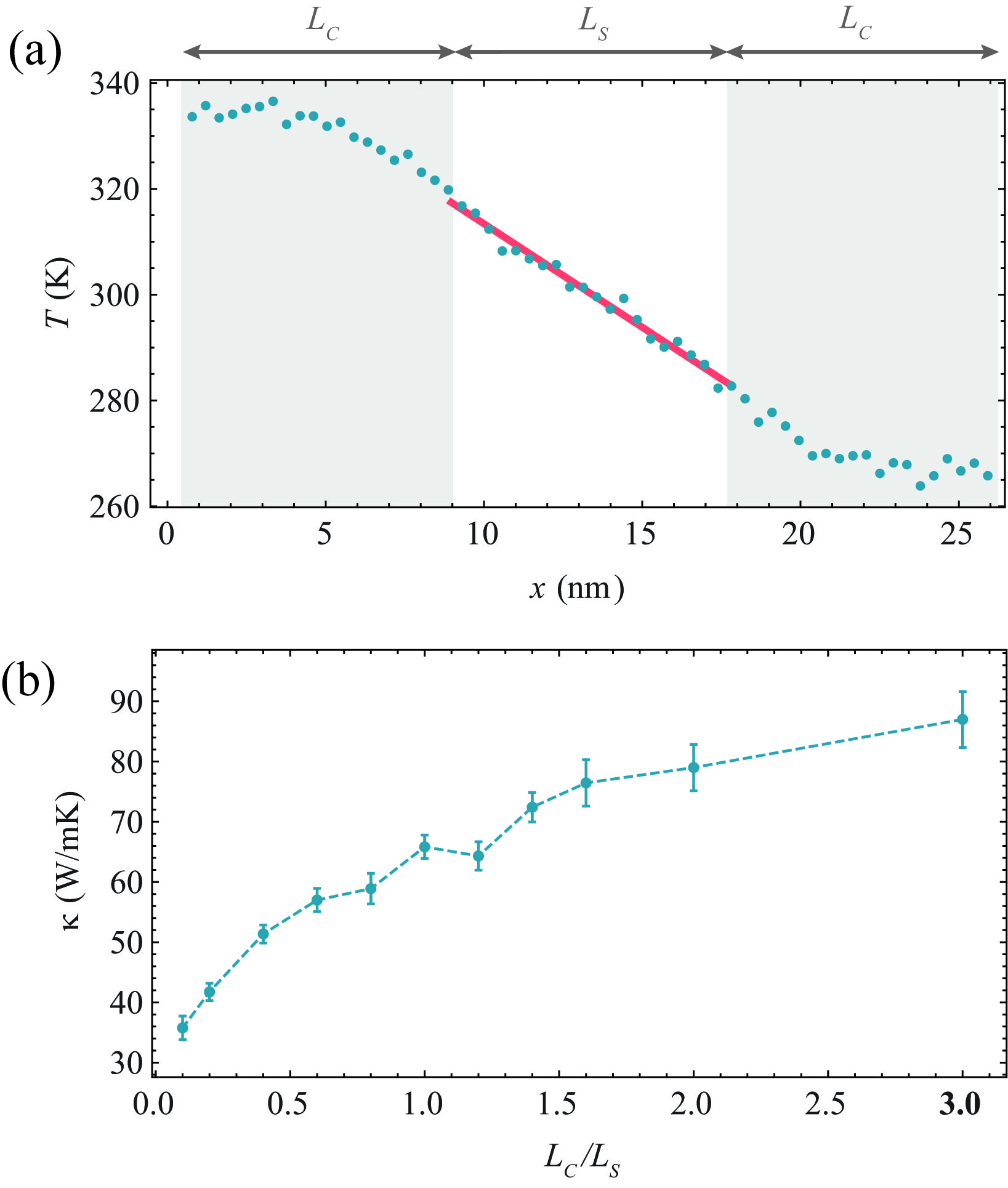}
\end{center}
\caption{(a)~Temperature profile and linear fit for an armchair graphene nanoribbon of width $W=\SI{1.2}{\nano\meter}$ and $L_S=L_C=\SI{8.4}{\nano\meter}$. The gray areas represent the contact regions of length $L_C$. (b)~Lattice thermal conductivity $\kappa$ as a function of the ratio $L_C/L_S$ for the above mentioned nanoribbons.}
\label{fig2}
\end{figure}

In this section we analyze the impact of contact sizes on our results. In our simulations we found that their size needs to be chosen carefully to obtain meaningful results. To do so, we consider an armchair GNR with fixed length $L_S$ and width $W$ and vary the contact length $L_C$ (see Figure~\ref{fig1} for a schematic view). We first plot the temperature profile across the system for a fixed value of $L_C$. In Figure~\ref{fig2}(a) one can observe that it is linear and smooth in the interface between the system and the contacts. 

Further we calculate $\kappa$ as a function of the contact length $L_C$, as displayed in Figure \ref{fig2}(b). We observe that $\kappa$ increases with $L_C$ and tends to saturate as $L_C/L_S$ becomes larger. Figure~\ref{fig2}(a) shows that the thermal gradient spans not only across the "device" part of the system (the central part of length $L_S$) but also across part of the contacts. Then the effective length of the system is larger than $L_S$ and varies with $L_C$. Therefore, the thermal conductivity is also contact size dependent because $\kappa$ is a length-dependent magnitude in nanometer-sized graphene nanoribbons \cite{Guo09,Park13}. This occurs because the contacts (heat source and heat sink) cannot be considered as isothermal classical boundary conditions. Although the average temperature remains constant, there is a temperature gradient within contacts. They are part of the system, so that vibrational modes are characterized by the whole dimension and not only by the size of the intermediate zone~\cite{Chantrenne04}. In order to avoid any dependence on the length dimension, the length of the system and the contacts (cold and hot regions) will be fixed throughout this work so that we take $L_C = \SI{12.6}{\nano\meter}$ and $L_S = \SI{25.4}{\nano\meter}$.

\section{Thermal conductivity of graphene nanorings}

\subsection{Role of dimensions}

\begin{figure}[tb]
\begin{center}
   \includegraphics[width=0.7\linewidth]{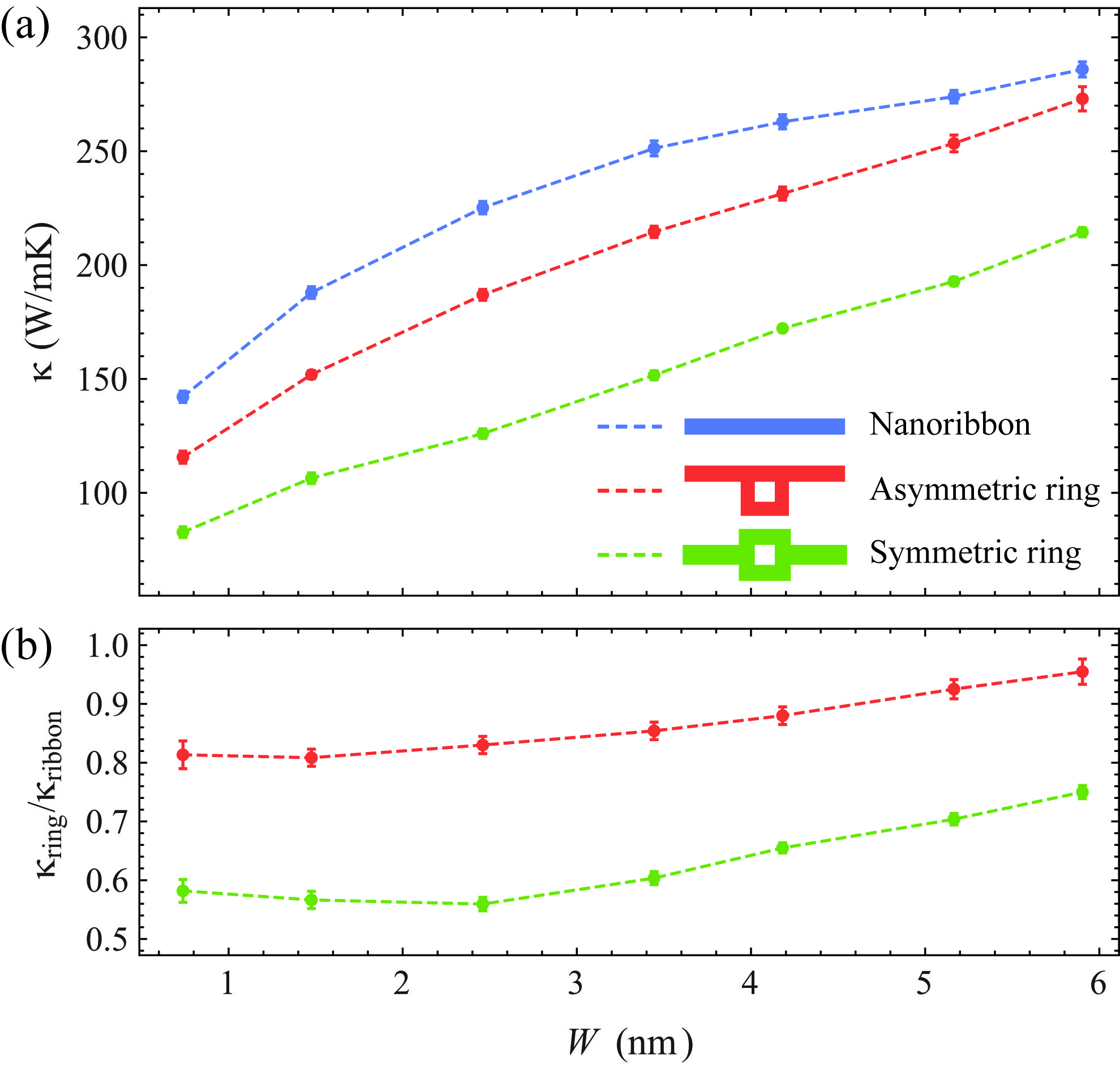}
\end{center}
\caption{(a)~Lattice thermal conductivity as a function of width $W$ for armchair ribbons and rings. (b)~Ratio $\kappa_\mathrm{ring}/\kappa_\mathrm{ribbon}$ as a function of the nanoribbon width $W$.}
\label{fig3}
\end{figure}

First, we fix $T=\SI{300}{\kelvin}$ and study how the thermal conductivity is affected by the width of the nanoribbons.  Figure~\ref{fig3}(a) shows the thermal conductivity for GNRs and two types of nanorings for widths $W$ up to $\SI{6}{\nano\meter}$. We refer to these as \emph{symmetric} or \emph{asymmetric} nanorings depending on the connection between the nanoring and the nanoribbons forming the leads (see Figure~\ref{fig3}). Our results show that $\kappa$ monotonously increases with the width $W$, both for nanoribbons and nanorings. This is in agreement with previous results~\cite{Guo09,Evans10}, where the same trend was found for armchair nanoribbons at room temperature. It can be understood as follows. Wider nanoribbons have a larger number of vibrational modes while the number of edge localized modes does not change. Thus, the edge effect decreases and $\kappa$ increases with $W$. At a threshold width, larger than the ones considered in this work, $\kappa$ will reach the value of graphene ($2000 - 4000\,$W/mK\cite{Pop12}) and then stay constant due to intermode scattering arising in the anharmonic lattice. Although $\kappa$ also increases with $W$ for nanorings, it remains lower than for nanoribbons at all $W$ considered by us. We interpret this as the effect of the mix of different edges, both armchair and zig-zag, that leads to a mismatch of the vibrational modes~\cite{Mazzamuto11} and by scattering at the bends~\cite{Li14,Yang13}. We note here that an introduction of more asymmetries in the rings, such as, different widths of the ring arms is expected to increase the scattering and reduce the conductivity even further.

Although absolute values of $\kappa$ obtained by NEMD simulations depend on the choice of interatomic potential, boundary conditions, simulated system dimensions and chosen method of imposing heat flux and temperature gradient, our results remain relevant because we are addressing the relative reduction of the thermal conductivity due to the nanostructuring.In the lower panel of Figure~\ref{fig3} the ratio $\kappa_\mathrm{ring}/\kappa_\mathrm{ribbon}$ is plotted, where $\kappa_\mathrm{ribbon}$ is the thermal conductivity of nanoribbons and $\kappa_\mathrm{ring}$ indicates the lattice thermal conductivity of the corresponding symmetric/asymmetric ring of the same nanoribbon width. Our results show that symmetric nanorings reduce the thermal conductivity more efficiently, reaching only~$60\%$ of the nanoribbon with the same width up to $\SI{3.5}{\nano\meter}$. This can be explained by the existence of more bends in symmetric rings, leading to a stronger scattering and suppression of the thermal conductivity. As the nanoribbon width decreases, the mismatch of vibration modes at different regions become more important and the suppression of $\kappa$ is stronger.

\subsection{Effect of temperature}

\begin{figure}[tb]
\begin{center}
   \includegraphics[width=0.7\linewidth]{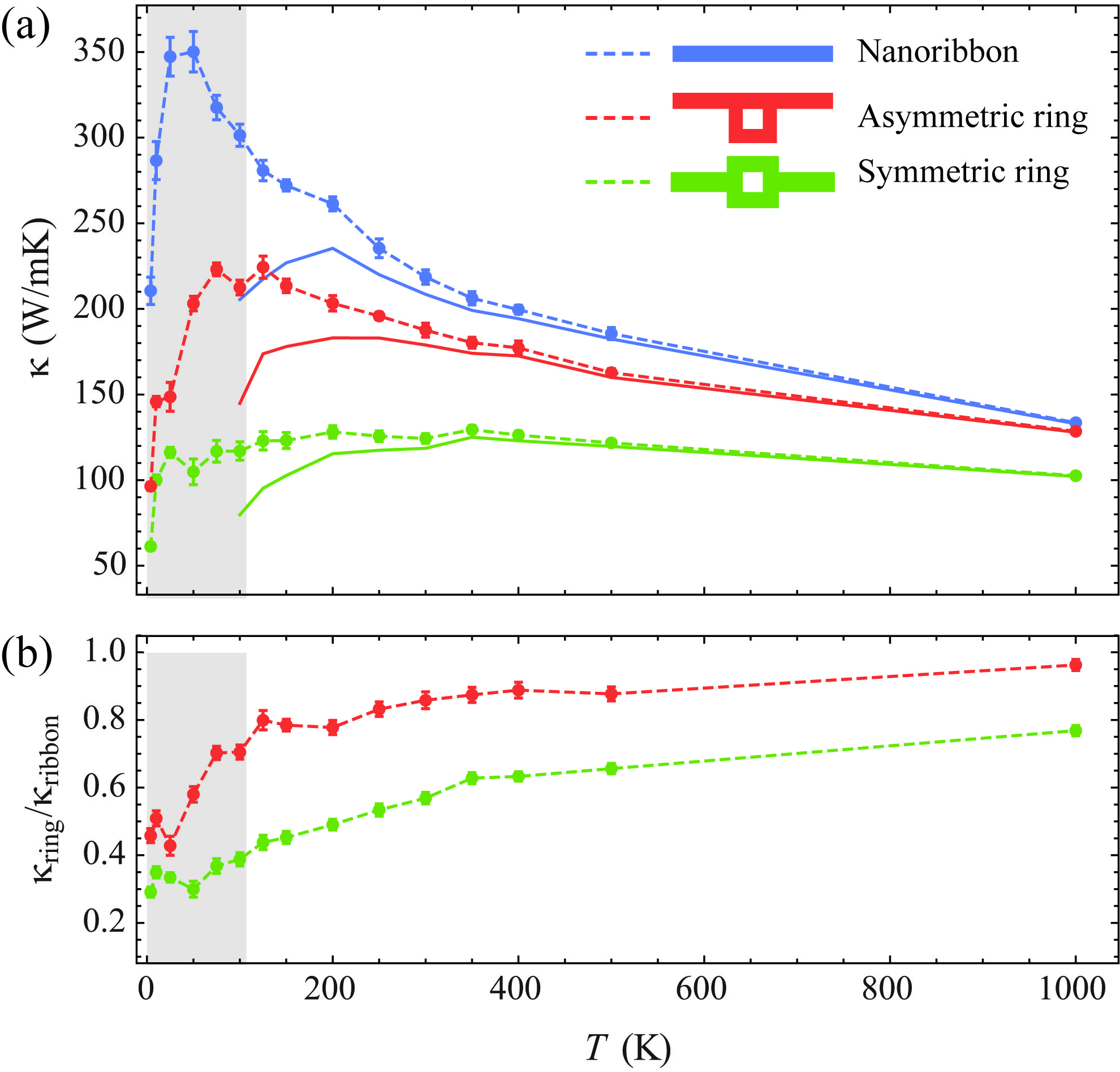}
\end{center}
\caption{(a)~Lattice thermal conductivity as a function of temperature for armchair nanoribbons and nanorings with $W=2.5$~nm. Solid lines represent the quantum corrected value of the conductivity for each structure. There is no quantum correction available for temperatures within the shadowed area. (b)~Ratio $\kappa_\mathrm{ring}/\kappa_\mathrm{ribbon}$ as a function of temperature $T$.}
\label{fig4}
\end{figure}

In this section we analyze the temperature dependence of $\kappa$. We take $W = \SI{2.5}{\nano\meter}$ as a typical value of the nanoribbon width and vary the temperature from $\SI{100}{\kelvin}$ up to $\SI{1000}{\kelvin}$. Classical NEMD simulations are considered valid near and above Debye's temperature ($T_\mathrm{D} \approx \SI{322}{\kelvin}$ for graphene nanoribbons \cite{Hu09}), where all vibrational modes are fully excited. At lower temperatures, quantum effects cannot be neglected. To mitigate this limitation, a quantum correction was developed using the scheme~\cite{Hu09, Hu09_2}
\begin{equation}
T_\mathrm{MD}=\frac{T_\mathrm{D}}{3}+\frac{2T_\mathrm{Q}^3}{T_\mathrm{D}^2}\int_{0}^{T_\mathrm{D}/T_\mathrm{Q}}\frac{x^2}{e^x-1}\,dx\ , 
\end{equation}
where $T_\mathrm{MD}$, $T_\mathrm{D}$ and $T_\mathrm{Q}$ are simulation temperature, Debye's temperature and quantum corrected temperature, respectively. The corrected thermal conductivity, $\kappa_\mathrm{D}=\kappa\, dT_\mathrm{MD}/dT_\mathrm{Q}$, is obtained by equating the heat fluxes obtained from Fourier's law in the classical (non-corrected) and quantum systems. This correction has been implemented in many works, also those studying graphene~\cite{Hu09,Hu09_2}. However, other works have questioned the validity of these quantum corrections~\cite{Turney09}. For this reason, in Figure~\ref{fig4} we plot both the quantum corrected and uncorrected values of the thermal conductivity. There is no quantum correction available at low temperatures (shadowed areas in Figure~\ref{fig4})~\cite{Hu09_2}. Our results show that for the three considered graphene structures, $\kappa$ first increases very quickly with $T$ until it reaches a maximum value and then it slowly decreases. When no quantum correction is considered, the maximum $\kappa$ value is reached at lower temperatures. We also find that low temperatures favor a reduction of the thermal conductivity in rings, both in symmetric and asymmetric configurations (see lower panel of Figure~\ref{fig4}). As before, the reduction is stronger for symmetric rings. For these rings, $\kappa$ is only about $40\%$ of the conductance of the corresponding nanoribbon for temperatures near $\SI{100}{\kelvin}$, and even for temperatures as high as $\SI{1000}{\kelvin}$ symmetric rings cause a significant decrease of the thermal conductivity ($\kappa_\mathrm{ring}<0.8\,\kappa_\mathrm{ribbon}$).

\subsection{Edge disorder and functionalization}

Thus far, we have considered ideal nanostructures while imperfections or disorder, can clearly affect the thermoelectric response. There exist different sources of disorder, such as charged impurities in the substrate, native defects and imperfections of the edges. The impact of the latter on the lattice thermal conductivity will probably be rather significant, especially in nanoscale systems. Following Refs.~\cite{Munarriz11,Munarriz12}, in order to estimate a possible impact of the edge disorder on the heat current, we consider disordered samples in which we delete carbon atoms from the edges. To do so, carbon atoms are randomly removed from the zigzag edges with some given probability $p$. To avoid dangling atoms in the armchair edges, pairs of neighbor atoms are removed with the same probability. We find that the thermal conductivity is almost independent of the exact position of the removed atoms as long as the probability $p$ remains the same. Even so, results are averaged over five realizations of the disordered sample. 

\begin{figure}[tb]
\begin{center}
   \includegraphics[width=0.7\linewidth]{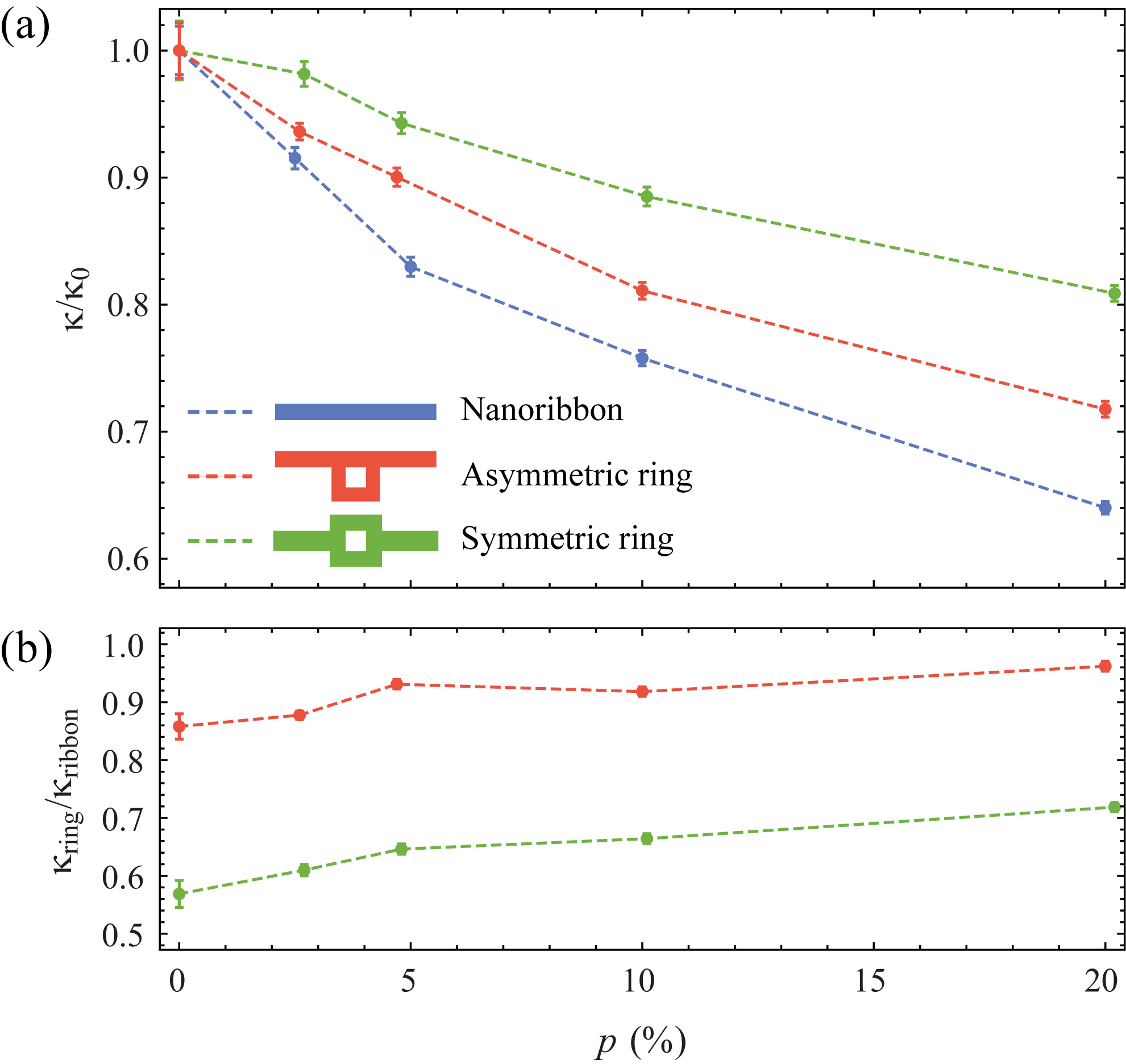}
\end{center}
\caption{(a)~Lattice thermal conductivity as a function of the probability of removal ($p$) for $W=2.5$~nm and $T=300$~K. All thermal conductivities are normalized by the thermal conductivity of the corresponding perfect ribbon or ring ($\kappa_0$). (b)~Ratio $\kappa_\mathrm{ring}/\kappa_\mathrm{ribbon}$ as a function of the probability of removal ($p$).}
\label{fig5}
\end{figure}

Figure~\ref{fig5}(a) shows the thermal conductivity of the nanoribbons and nanorings as a function of the probability $p$ when $W = \SI{2.5}{\nano\meter}$ and $\SI{300}{\kelvin}$. All thermal conductivities are normalized to the thermal conductivity of the corresponding perfect nanostructure, denoted by $\kappa_0$. Our results show that rough edges decrease $\kappa$ in the three studied nanostructures. This is in good agreement with the results reported in Refs.~\cite{Savin10,Evans10}, where it was found that rough edges can cause a suppression of the thermal conductivity in nanoribbons due to the scattering of vibrational modes. Furthermore, Figure \ref{fig5}(a) shows nanoribbons would be the most affected, while symmetric rings would be the least. It is worth mentioning that $\kappa_\mathrm{ring}(p) < \kappa_\mathrm{ribbon}(p)$ continues to be true in spite of the already marked decrease observed for $\kappa_\mathrm{ribbon}(p)$ when increasing $p$. Figure \ref{fig5}(b) demonstrates that the thermal conductivity in symmetric rings is about $70\%$ of the corresponding ribbon conductivity for $p=20\%$ (which is the largest probability of removal considered in this work).

One major issue regarding heat transport in narrow GNRs is to elucidate up to what extent would chemical functionalization affect the results. Among various functional groups, hydrogen has attracted considerable interest in recent years (see Ref.~\cite{Pei11} and references therein). In order to answer this question, we have conducted further NEMD simulations of GNRs and rings with edges saturated by hydrogen atoms. We found that the thermal conductivity slightly decreases as compared to open GNRs and rings of the same size. This finding seems consistent with NEMD simulations carried out in hydrogenated GNRs~\cite{Pei11} and GNRs with hydrogen termination~\cite{Evans10}. The decrease of the thermal conductivity is larger in rings than in GNRs because the larger edge length of the formers. However, in both cases the reduction of the thermal conductivity is not significant and the effects of hydrogen-saturated edges can be safely neglected.

\section{Conclusions}

We have studied the lattice thermal conductivity of graphene nanorings and nanoribbons by means of NEMD. We found a significant reduction of the thermal conductivity $\kappa_\mathrm{lat}$ in symmetric rings, especially for narrow widths and low temperatures, as compared to uniform GNRs. Even at temperatures as high as $\SI{1000}{\kelvin}$, our results show a substantial decrease of $\kappa_\mathrm{lat}$. The impact of rough edges on heat transport was also addressed and we concluded that it is higher in nanoribbons. Nevertheless, the thermal conductivity of disordered nanorings is considerably smaller than that of nanoribbons at the same magnitude of disorder. Therefore, nanorings present two main advantages for exploiting their thermoelectric properties as compared to nanoribbons. First, quantum interference effects enhance the ratio $\sigma/\kappa_\mathrm{el}$ and make it much larger than the universal value predicted by the Wiedemann-Franz law in ohmic metals. Second, the scattering of vibrational modes at bends yields a strong reduction of the lattice thermal conductivity $\kappa_\mathrm{lat}$, which can probably be further decreased in the case of more irregular rings having arms of different widths or lengths.

\section*{Acknowledgments}

The authors are grateful to R.\ Brito for helpful discussions. M.\ S.-B.\ thanks the Theoretical Physics Group of the  University of Warwick for the warm hospitality. Work at Madrid has been supported by MINECO under grants MAT2013-46308, MAT2016-75955 and MAT2016-63955-R. Calculations were performed at the \textit{Cluster de C\'{a}lculo de Alta Capacidad para T\'{e}cnicas F\'{\i}sicas}, funded by the Universidad Complutense and the EU under the FEDER  program. UK research data statement: All data accompanying this publication are directly available within the publication.

\section*{References}

\bibliography{references}

\begin{thebibliography}{10}
\expandafter\ifx\csname url\endcsname\relax
  \def\url#1{\texttt{#1}}\fi
\expandafter\ifx\csname urlprefix\endcsname\relax\def\urlprefix{URL }\fi
\expandafter\ifx\csname href\endcsname\relax
  \def\href#1#2{#2} \def\path#1{#1}\fi

\bibitem{Pop06}
E.~Pop, S.~Sinha, K.~E. Goodson, Heat generation and transport in
  nanometer-scale transistors, Proc. IEEE 94 (2006) 1587.

\bibitem{Vasileska08}
K.~Vasileska, D.and~Raleva, S.~M. Goodnick, Modeling heating effects in
  nanoscale devices: the present and the future, J. Comput. Electron. 7 (2008)
  66.

\bibitem{Pop10}
E.~Pop, Energy dissipation and transport in nanoscale devices, Nano Res. 3
  (2010) 147.

\bibitem{Goldsmid10}
H.~J. Goldsmid, Introduction to thermoelectricity, Springer, Berlin, 2010.

\bibitem{Villagonzalo1999}
C.~Villagonzalo, R.~R{\"{o}}mer, M.~Schreiber, {Thermoelectric transport
  properties in disordered systems near the {A}nderson transition}, Eur. Phys.
  J. B 12.

\bibitem{Franz1853}
R.~Franz, G.~Wiedemann, {Ueber die {W}{\"{a}}rme-Leitungsf{\"{a}}higkeit der
  {M}etalle}, Annalen der Physik und Chemie 165~(8) (1853) 497.

\bibitem{Hicks93}
L.~D. Hicks, M.~S. Dresselhaus, Thermoelectric figure of merit of a
  one-dimensional conductor, Phys. Rev. B 47 (1993) 16631.

\bibitem{Khitun00}
A.~Khitun, A.~Balandin, J.~L. Liu, K.~L. Wang, In-plane lattice thermal
  conductivity of a quantum-dot superlattice, J. Appl. Phys. 88 (2000) 696.

\bibitem{Balandin03}
A.~A. Balandin, O.~L. Lazarenkova, Mechanism for thermoelectric figure-of-merit
  enhancement in regimented quantum dot superlattices, Appl. Phys. Lett. 82
  (2003) 415.

\bibitem{Sadeghi15}
H.~Sadeghi, S.~Sangtarash, C.~J. Lambert, Enhancing the thermoelectric figure
  of merit in engineered graphene nanoribbons, Beilstein J. Nanotechnol. 6
  (2015) 1176.

\bibitem{Venkata01}
R.~Venkatasubramanian, E.~Siivola, T.~Colpitts, B.~O'Quinn, Thin-film
  thermoelectric devices with high room-temperature figures of merit, Nature
  (London) 413 (2001) 597.

\bibitem{Harman02}
T.~C. Harman, P.~J. Taylor, M.~P. Walsh, B.~E. LaForge, Quantum dot
  superlattice thermoelectric materials and devices, Science 297 (2002) 2229.

\bibitem{Hochbaum08}
A.~I. Hochbaum, R.~Chen, R.~D. Delgado, W.~Liang, E.~C. Garnett, M.~Najarian,
  A.~Majumdar, P.~Yan, Enhanced thermoelectric performance of rough silicon
  nanowires, Nature (London) 451 (2008) 163.

\bibitem{Boukai08}
A.~I. Boukai, Y.~Bunimovich, J.~Tahir-Kheli, J.-K. Yu, W.~A. Goddard, J.~R.
  Heath, Silicon nanowires as efficient thermoelectric materials, Nature
  (London) 451 (2008) 168.

\bibitem{Mahan96}
G.~D. Mahan, J.~O. Sofo, The best thermoelectric, Proc. Natl. Acad. Sci. USA 93
  (1996) 7436.

\bibitem{GomezSilva12}
G.~G\'omez-Silva, O.~\'Avalos-Ovando, M.~L.~L. de~Guevara, P.~A. Orellana,
  Enhancement of thermoelectric efficiency and violation of the
  {W}iedemann-{F}ranz law due to {F}ano effect, J. Appl. Phys. 111 (2012)
  053704.

\bibitem{Zheng12}
J.~Zheng, M.-J. Zhu, F.~Chi, Fano effect on the thermoelectric efficiency in
  parallel-coupled double quantum dots, J. Low Temp. Phys. 166 (2012) 208.

\bibitem{Garcia13}
V.~M. Garc\'{\i}a-Su\'arez, R.~Ferrad\'as, J.~Ferrer, Impact of {F}ano and
  {B}reit-{W}igner resonances in the thermoelectric properties of nanoscale
  junctions, Phys. Rev. B 88 (2013) 235417.

\bibitem{Fu15}
H.-H. Fu, L.~Gu, D.-D. Wu, Z.-Q. Zhang, Enhancement of the thermoelectric
  figure of merit in {DNA}-like systems induced by {F}ano and {D}icke effects,
  Phys. Chem. Chem. Phys. 17 (2015) 11077.

\bibitem{SaizBretin16}
M.~A. Sierra, M.~Saiz-Bret\'{\i}n, F.~Dom\'{\i}nguez-Adame, D.~S\'anchez,
  Interactions and thermoelectric effects in a parallel-coupled double quantum
  dot, Phys. Rev. B 93 (2016) 235452.

\bibitem{Wang16}
R.-N. Wang, G.-Y. Dong, S.-F. Wang, G.-S. Fu, J.-L. Wang, Impact of contact
  couplings on thermoelectric properties of anti, {F}ano, and {B}reit-{W}igner
  resonant junctions, J. Appl. Phys. 120 (2016) 184303.

\bibitem{Koumoto13}
K.~Koumoto, T.~Mori, Thermoelectric nanomaterials. {M}aterials design and
  applications, Springer, Berlin, 2013.

\bibitem{Baringhaus14}
J.~Baringhaus, M.~Ruan, F.~Edler, A.~Tejeda, M.~Sicot, A.~Taleb-Ibrahimi, A.-P.
  Li, Z.~Jiang, E.~H. Conrad, C.~Berger, C.~Tegenkamp, W.~A. de~Heer,
  Exceptional ballistic transport in epitaxial graphene nanoribbons, Nature
  (London) 506 (2014) 349.

\bibitem{Russo08}
S.~Russo, J.~B. Oostinga, D.~Wehenkel, H.~B. Heersche, S.~S. Sobhani, L.~M.~K.
  Vandersypen, A.~F. Morpurgo, Observation of {A}haronov-{B}ohm conductance
  oscillations in a graphene ring, Phys. Rev. B 77 (2008) 085413.

\bibitem{Smirnov12}
D.~Smirnov, H.~Schmidt, R.~J. Haug, Aharonov-{B}ohm effect in an electron-hole
  graphene ring system, Appl. Phys. Lett. 100 (2012) 203114.

\bibitem{Schelter12}
J.~Schelter, P.~Recher, B.~Trauzettel, The {A}haronov-{B}ohm effect in graphene
  rings, Solid State Commun. 152 (2012) 1411.

\bibitem{Cabosart14}
D.~Cabosart, S.~Faniel, F.~Martins, B.~Brun, A.~Felten, V.~Bayot, B.~Hackens,
  Imaging coherent transport in a mesoscopic graphene ring, Phys. Rev. B 90
  (2014) 205433.

\bibitem{Samal15}
M.~Samal, N.~Barange, D.-H. Ko, K.~Yun, Graphene quantum rings doped
  {PEDOT:PSS} based composite layer for efficient performance of optoelectronic
  devices, J. Phys. Chem. 119  19619.

\bibitem{SaizBretin15}
M.~Saiz-Bret\'in, A.~V. Malyshev, P.~A. Orellana, F.~Dom\'inguez-Adame,
  Enhancing thermoelectric properties of graphene quantum rings, Phys. Rev. B
  91 (2015) 085431.

\bibitem{Savin10}
A.~V. Savin, Y.~S. Kivshar, B.~Hu, Suppression of thermal conductivity in
  graphene nanoribbons with rough edges, Phys. Rev. B 82 (2010) 195422.

\bibitem{Hu10}
J.~Hu, S.~Schiffli, A.~Vallabhaneni, X.~Ruan, Y.~P. Chen, Tuning the thermal
  conductivity of graphene nanoribbons by edge passivation and isotope
  engineering: {A} molecular dynamics study, Appl. Phys. Lett. 97 (2010)
  133107.

\bibitem{Mazzamuto11}
F.~Mazzamuto, V.~Hung~Nguyen, Y.~Apertet, C.~Caër, C.~Chassat,
  J.~Saint-Martin, P.~Dollfus, Enhanced thermoelectric properties in graphene
  nanoribbons by resonant tunneling of electrons, Phys. Rev. B 83 (2011)
  235426.

\bibitem{Li14}
K.-M. Li, Z.-X. Xie, K.-L. Su, W.-H. Luo, Y.~Zhang, Ballistic thermoelectric
  properties in double-bend graphene nanoribbons, Phys. Lett. A 378 (2014)
  1383.

\bibitem{Zhang12}
H.-S. Zhang, Z.-X. Guo, X.-G. Gong, J.-X. Cao, Thermal conductivity of
  sawtooth-like graphene ribbons: {A} molecular dynamics study, J. Appl. Phys.
  112 (2012) 123508.

\bibitem{Chen10}
Y.~Chen, T.~Jayasekera, A.~Calzolari, K.~W. Kim, M.~Buongiorno~Nardelli,
  Thermoelectric properties of graphene nanoribbons, junctions and
  superlattices, J. Phys.: Condens. Mat. 22 (2010) 372202.

\bibitem{Xu10}
Y.~Xu, X.~Chen, J.-S. Wang, B.-L. Gu, W.~Duan, Thermal transport in graphene
  junctions and quantum dots, Phys. Rev. B 81 (2010) 195425.

\bibitem{Wu10}
Z.~Wu, Z.~Z. Zhang, K.~Chang, F.~M. Peeters, Quantum tunneling through graphene
  nanorings, Nanotechnology 21 (2010) 185201.

\bibitem{Munarriz11}
J.~Mun{\'a}rriz, F.~Dom{\'i}nguez-Adame, A.~V. Malyshev, Towards graphene-based
  quantum interference devices, Nanotechnology 22 (2011) 365201.

\bibitem{Mrenka16}
A.~Mre\ifmmode \acute{n}\else \'{n}\fi{}ca-Kolasi\ifmmode~\acute{n}\else
  \'{n}\fi{}ska, B.~Szafran, Lorentz force effects for graphene
  {A}haronov-{B}ohm interferometers, Phys. Rev. B 94 (2016) 195315.

\bibitem{Sousa17}
D.~J.~P. de~Sousa, A.~Chaves, J.~M. PereiraJr., G.~A. Farias, Interferometry of
  {K}lein tunnelling electrons in graphene quantum rings, J. Appl. Phys. 121
  (2017) 024302.

\bibitem{Munarriz12}
J.~Mun{\'a}rriz, F.~Dom{\'i}nguez-Adame, P.~A. Orellana, A.~V. Malyshev,
  Graphene nanoring as a tunable source of polarized electrons, Nanotechnology
  23 (2012) 205202.

\bibitem{Farghadan13}
R.~Farghadan, A.~Saffarzadeh, E.~H. Semiromi, Magnetic edge states in
  {A}haronov-{B}ohm graphene quantum rings, J. Appl. Phys. 114~(21) (2013)
  214314.

\bibitem{LAMMPS}
\url{http://lammps.sandia.gov}.

\bibitem{Guo09}
Z.~Guo, D.~Zhang, X.-G. Gong, Thermal conductivity of graphene nanoribbons,
  Appl. Phys. Lett. 95 (2009) 163103.

\bibitem{Hu09}
J.~Hu, R.~X., Y.-P. Chen, Thermal conductivity and thermal rectification in
  graphene nanoribbons: A molecular dynamics study, Nano Lett. 9 (2009) 2730.

\bibitem{Evans10}
W.~J. Evans, L.~Hu, P.~Keblinski, Thermal conductivity of graphene ribbons from
  equilibrium molecular dynamics: effect of ribbon width, edge roughness, and
  hydrogen termination, Appl. Phys. Lett. 96 (2010) 203112.

\bibitem{Plimpton95}
S.~Plimpton, Fast parallel algorithms for short-range molecular dynamics, J.
  Comput. Phys. 117 (1995) 1.

\bibitem{Frenkel96}
D.~Frenkel, B.~Smit, Understanding molecular simulation: {F}rom algorithms to
  applications, Academic Press, London, 2002.

\bibitem{Lukes00}
J.~R. Lukes, D.~Y. Li, X.-G. Liang, C.-L. Tien, Molecular dynamics study of
  solid thin-film thermal conductivity, J. Heat Transf. 122 (2000) 536.

\bibitem{Pei11}
Q.-X. Pei, Z.-D. Sha, Y.-W. Zhang, A theoretical analysis of the thermal
  conductivity of hydrogenated graphene, Carbon 49 (2011) 4752.

\bibitem{Kaiser17}
J.~Kaiser, T.~Feng, J.~Maassen, X.~Wang, X.~Ruan, M.~Lundstrom, Thermal
  transport at the nanoscale: {A} {F}ourier's law vs. phonon {B}oltzmann
  equation study, J. Appl. Phys. 121 (2017) 044302.

\bibitem{Stuart00}
S.~J. Stuart, A.~B. Tutein, J.~A. Harrison, A reactive potential for
  hydrocarbons with intermolecular interactions, J. Chem. Phys. 112 (2000)
  6472.

\bibitem{Ng12}
T.~Y. Ng, J.~J. Yeo, Z.~S. Liu, A molecular dynamics study of the thermal
  conductivity of graphene nanoribbons containing dispersed
  {S}tone-{T}hrower-{W}ales defects, Carbon 50 (2012) 4887.

\bibitem{Yang13}
P.~Yang, Y.~Tang, H.~Yang, J.~Gong, Y.~Liu, Y.~Zhao, X.~Yu, Thermal management
  performance of bent graphene nanoribbons, RSC Advances 3 (2013) 17349.

\bibitem{Shiomi14}
J.~Shiomi, Nonequilibrium molecular dynamics methods for lattice heat
  conduction calculations, Annu. Rev. Heat Transf. 17 (2014) 177.

\bibitem{Park13}
M.~Park, S.-C. Lee, Y.-S. Kim, Length-dependent lattice thermal conductivity of
  graphene and its macroscopic limit, J. Appl. Phys. 114 (2013) 053506.

\bibitem{Chantrenne04}
P.~Chantrenne, J.-L. Barrat, Finite size effects in determination of thermal
  conductivities: {C}omparing molecular dynamics results with simple models, J.
  Heat Transf. 126(4) (2004) 577.

\bibitem{Pop12}
E.~Pop, V.~Varshney, A.~K. Roy, Thermal properties of graphene: {F}undamentals
  and applications, MRS Bulletin 37 (2012) 1273.

\bibitem{Hu09_2}
J.~Hu, X.~Ruan, Z.~Jiang, Y.~P. Chen, Molecular dynamics calculation of thermal
  conductivity of graphene nanoribbons, AIP Conference Proceedings 1173 (2009)
  135.

\bibitem{Turney09}
J.~E. Turney, J.~H. McGaughey, C.~H. Amon, Assessing the applicability of
  quantum corrections to classical thermal conductivity predictions, Phys. Rev.
  B 79 (2009) 224305.

\end{thebibliography}

\end{document}